\documentclass[prl,twocolumn,showpacs]{revtex4}

\usepackage{amsmath}
\usepackage{graphicx}
\usepackage{dcolumn}
\usepackage{bm}

\begin{document}

\preprint{ITP/UU-XXX}

\title{Lifshitz Point in the Phase Diagram of Resonantly Interacting ${}^{6}$Li-${}^{40}$K Mixtures}

\author{K. B. Gubbels}
\email{K.B.Gubbels@uu.nl}
\author{J. E. Baarsma}
\author{H. T. C. Stoof}

\affiliation{
Institute for Theoretical Physics, Utrecht University,\\
Leuvenlaan 4, 3584 CE Utrecht, The Netherlands}


\begin{abstract}

We consider a strongly interacting ${}^{6}$Li-${}^{40}$K mixture,
which is imbalanced both in the masses and the densities of the
two fermionic species. At present, it is the experimentalist's
favorite for reaching the superfluid regime. We construct an
effective thermodynamic potential that leads to excellent
agreement with Monte Carlo results for the normal state. We use it
to determine the universal phase diagram of the mixture in the
unitarity limit, where we find, in contrast to the mass-balanced
case, the presence of a Lifshitz point. This point is
characterized by the effective mass of the Cooper pairs becoming
negative, which signals an instability towards a supersolid phase.
\end{abstract}

\pacs{03.75.-b, 67.40.-w, 39.25.+k}

\maketitle

\emph{Introduction.} --- Ultracold atomic Fermi mixtures are
presently at the center of attention of both experimental and
theoretical physicists. Due to the amazing control over this
system many fundamental discoveries have been made, while more are
likely to follow soon. These discoveries started with reaching the
so-called BEC regime where the balanced two-component Fermi
mixture turns superfluid due to the Bose-Einstein condensation
(BEC) of molecules \cite{Greiner,Jochim}. By using a Feshbach
resonance to vary the interaction strength between atoms in
different spin states, the BEC-BCS crossover between a
Bose-Einstein condensate of molecules and a BCS superfluid of
Cooper pairs could be directly observed \cite{Regal}. Since
pairing is optimal for an equal amount of atoms in each spin state
and is absent for the noninteracting fully polarized system, a
phase transition occurs as a function of spin imbalance
\cite{Ketterle,Hulet}. The phase diagram of the polarized mixture
was for strong interactions found to be governed by a tricritical
point that resulted in the observation of phase separation
\cite{Hulet,Shin}. The presence of gapless Sarma superfluidity is
also likely to be present in this system \cite{Gubbels,Diederix},
but has not been unambiguously identified yet.

Most recently, experiments have indicated that the physical
consequences of yet another parameter can be explored, namely that
of a mass imbalance between the two fermionic components. A very
promising mixture in this respect consists of ${}^6$Li and
${}^{40}$K, which has a mass ratio of 6.7. Several accessible
Feshbach resonances are identified in the mixture \cite{Walraven},
while both species have also been simultaneously cooled into the
degenerate regime \cite{Dieckmann}. So far, experimental interest
has focused on the BEC side of the Feshbach resonance, where
molecules are formed. Although the mass imbalance itself does not
lead to fundamental new physics here, this situation changes when
the heteronuclear molecules are optically pumped to their ground
state \cite{Ye}. Then the molecules acquire a large electric
dipole moment, which gives rise to anisotropic long-ranged
interactions. In an optical lattice, this can lead to supersolid
phases \cite{Lewenstein}.

In this Letter we focus on a different regime, namely the
so-called unitarity limit, where the $s$-wave scattering length of
the interspecies interaction diverges. Here, the size of the
Cooper pairs is comparable to the average interparticle distance
and the pairing is a many-body effect. The mass imbalance has a
profound effect on the pairing now, because it strongly alters the
Fermi spheres. We show that for the sufficiently large mass ratio
of the ${}^6$Li-${}^{40}$K mixture, the phase diagram not only
encompasses all the exciting physics known from the mass-balanced
case, but is even much richer. Similar to the solely
spin-imbalanced case is the presence of phase separation
\cite{Parish}, which can occur due to the mismatch of the Fermi
surfaces. Also similar is that gapless Sarma superfluidity is
unstable at zero temperature \cite{Parish}, while there is a
predicted crossover to the Sarma phase at nonzero temperatures
\cite{Gubbels}. However, the most exciting difference that we find
is the presence of a Lifshitz point in the phase diagram.

At a Lifshitz point the transition to the superfluid phase
undergoes a dramatic change of character. Rather than preferring a
homogeneous order parameter, the system now becomes an
inhomogeneous superfluid. This exotic possibility was early
investigated for the weakly interacting mass-balanced case by
Larkin and Ovchinnikov (LO), who considered a superfluid with a
single standing-wave order parameter \cite{Larkin}. This is
energetically more favorable than the plane-wave case studied by
Fulde and Ferrell (FF) \cite{Fulde}. Since the LO phase results in
periodic modulations of the particle densities, it is a supersolid
\cite{Bulgac}. The FF and LO phases have intrigued the physics
community for many decades, but so far remained elusive in
experiments with atomic Fermi mixtures. Typically, Lifshitz points
are predicted at weak interactions where the critical temperatures
are very low. However, in this Letter we show that the very
special phase diagram of the ${}^6$Li-${}^{40}$K mixture contains
both a Lifshitz and a tricritical point in the unitarity limit, as
shown in Fig. 1. This is in sharp contrast to the mass-balanced
case, where at unitarity a large body of theory only finds a
tricritical point, in agreement with experiments \cite{Shin}.

\begin{figure*}[t!]
\includegraphics[width=1.8\columnwidth]{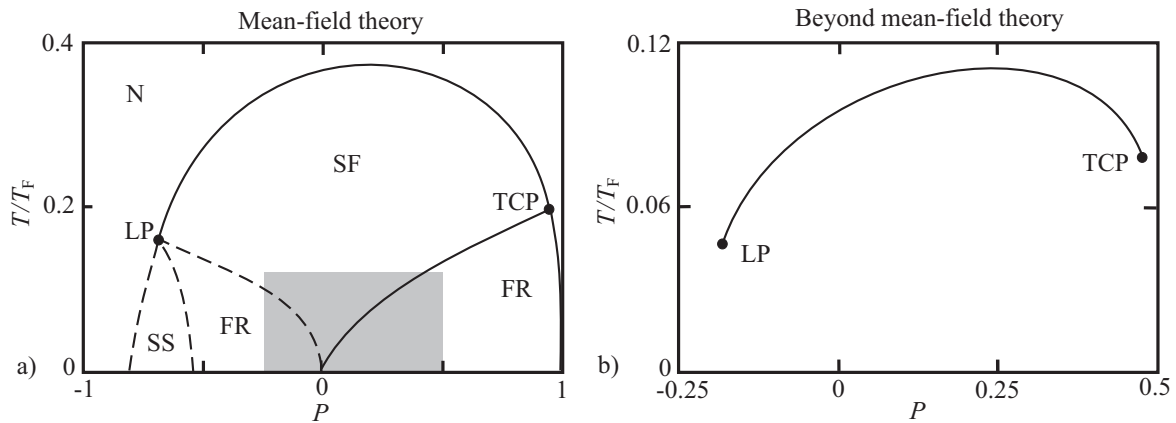}
\caption{\label{phasediagram} Universal phase diagram for the
homogeneous ${}^6$Li-${}^{40}$K mixture in the unitarity limit as
a function of temperature $T$ and polarization $P$. The
temperature is scaled with the reduced Fermi temperature $k_{\rm
B} T_{\rm F}=\epsilon_{\rm F}=\hbar^2 (3 \pi^2 n)^{2/3}/2m$, where
$m$ is twice the reduced mass and $n$ is the total particle
density. The result of mean-field theory is shown in panel a). For
a majority of light ${}^6$Li atoms there is a tricritical point
(TCP), at which the normal state (N), the homogeneous superfluid
state (SF), and the forbidden region (FR) meet each other. For a
majority of heavy ${}^{40}$K atoms there is a Lifshitz point (LP),
where there is an instability towards supersolidity (SS). The size
of the supersolid stability region is not calculated within our
theory and the dashed lines are therefore only guides to the eye.
The grey area sets the scale for panel b), where fluctuation
effects are taken into account to calculate the location of the
Lifshitz and the tricritical point more accurately.}
\end{figure*}

In first instance, all these expectations follow from mean-field
theory, which is useful for a qualitative description of the
physics. However, it cannot be trusted quantitatively in the
unitary regime, since it vastly overestimates critical
temperatures in that case. It is thus important to understand how
much these are lowered by fluctuation effects. In the
mass-balanced case two effects are dominant, namely the fermionic
selfenergies and the screening of the interaction due to
particle-hole fluctuations \cite{Koos}. Taking these into account
gives in an intuitive manner results that compare well with Monte
Carlo calculations \cite{Burovski} and experiment \cite{Shin}. We
find that an extension of this procedure to the
${}^6$Li-${}^{40}$K mixture leads to a reduction of the mean-field
critical temperatures by an experimentally relevant factor of 3.
This makes it more difficult, but not impossible, to reach the
superfluid regime. Moreover, the presence of the Lifshitz point in
the phase diagram is unaffected by the fluctuations, allowing for
an experimental study of supersolidity in the ${}^6$Li-${}^{40}$K
mixture.

\emph{Phase diagram.} --- The critical properties of the
superfluid transition in a fermionic mixture are determined by the
effective Landau theory for the superfluid order parameter $\Delta
({\bf x})$. Although we consider no external potential, the order
parameter may still vary in space due to a spontaneous breaking of
translational symmetry. Close to the continuous superfluid
transition, we expand the exact effective thermodynamic potential
as
\begin{eqnarray}\label{omlan}
\Omega[\Delta] &=&\int d {\bf x}\left\{\gamma|\nabla \Delta|^2
+\alpha| \Delta|^2+\frac{\beta}{2}|\Delta |^4+...\right\},
\end{eqnarray}
where the challenge is to express the expansion parameters in
terms of the temperature $T$ and the atomic chemical potentials
$\mu_{\pm}$. The upper (lower) sign refers to the light ${}^6$Li
(heavy ${}^{40}$K) atoms. A phase transition has occurred when the
global minimum of $\Omega$ is located at a nonzero order parameter
$\langle \Delta({\bf x}) \rangle$, which describes a condensate of
pairs. When $\gamma(T)$ is positive, the pairs have a positive
effective mass and their center-of-mass state of lowest energy is
at zero momentum. Then, we can consider a homogeneous pairing
field $\Delta$, for which a second-order transition occurs at a
critical temperature $T_c$ determined by $\alpha (T_{\rm c })=0$.

A continuous transition only occurs when the minimum at small
values of $\langle\Delta\rangle$ is global, which is not
necessarily the case. The expansion of $\Omega$ may contain higher
powers of $|\Delta|^2$ that have negative coefficients, leading to
a first-order transition with a jump in the order parameter when
$\Omega[0]=\Omega[\langle \Delta \rangle]$. Second-order behavior
turns into first-order behavior when $\beta(T)$ becomes negative,
so that the temperature $T_{\rm c3}$ at the tricritical point
(TCP) is determined by $\alpha(T_{\rm c3})=0$ and $\beta(T_{\rm
c3})=0$. Another intriguing possibility is that not $\beta(T)$,
but rather $\gamma(T)$ goes to zero. This leads to a Lifshitz
point (LP), which is thus determined by $\alpha(T_{\rm L})=0$ and
$\gamma(T_{\rm L})=0$. Since the effective mass of the Cooper
pairs becomes negative below the Lifshitz point, it is
energetically favorable for them to have kinetic energy and form a
superfluid at nonzero momentum. This can be established in many
ways, namely through a standing wave \cite{Larkin} or a more
complicated superposition of plane waves \cite{Yip,Mora}. Due to
the variety of possibilities it is hard to predict which lattice
structure is most favorable. However, the fact that they all
emerge from the Lifhitz point facilitates the experimental search
for supersolidity in the ${}^6$Li-${}^{40}$K mixture. Moreover, a
nonzero gap $\langle\Delta\rangle$ gives rise to a sizeable
superfluid fraction, showing that the LO-like incommensurate
supersolid is very different from the commensurate supersolid
studied in ${}^4$He.

We continue our discussion at the mean-field level, which is
useful for further explaining the relevant physics. Starting point
is the BCS thermodynamic potential density for the
${}^6$Li-${}^{40}$K mixture with masses $m_{\pm}$ \cite{Parish},
\begin{eqnarray} \label{thermpot}
\omega_{\rm BCS}[\Delta;\mu_{\sigma}]&=&\int\frac{d{\bf
k}}{(2\pi)^3}\bigg\{ \epsilon(\mathbf{k}) -\mu - \hbar
\omega(\mathbf{k})+\frac{|\Delta|^2}{2 \epsilon(\mathbf{k})}
\nonumber \\
&-&k_{\rm B}T\sum_{\sigma=\pm}\ln\left(1+e^{-
\hbar\omega_{\sigma}(\mathbf{k})/k_{\rm B}T}\right)\bigg\},
\end{eqnarray}
where we introduced the average chemical potential
$\mu=(\mu_{+}+\mu_{-})/2$, the difference $h=(\mu_{+}-\mu_{-})/2$,
and the reduced kinetic energy $\epsilon(\mathbf{k})=\hbar^2
\mathbf{k}^2/2m$ with $m=2 m_+ m_-/(m_++m_-)$. The Bogoliubov
quasiparticle dispersions become
$\hbar\omega_{\sigma}(\mathbf{k})= \hbar\omega(\mathbf{k}) -
\sigma[2h - \epsilon_+(\mathbf{k})+\epsilon_-(\mathbf{k})]/2$ with
$\hbar\omega({\mathbf{k}})=
\sqrt{[\epsilon({\mathbf{k}})-\mu]^2+|\Delta|^2}$ and
$\epsilon_{\sigma}({\mathbf{k}})=\hbar^2
\mathbf{k}^2/2m_{\sigma}$. We can now apply the exact critical
conditions to $\omega_{\rm BCS}$ and obtain the mean-field phase
diagram. Although the BCS potential neglects fluctuations in the
order parameter, it is expected that fluctuation effects only
result in quantitative corrections. This expectation stems from
the strongly interacting experiments for the mass-balanced case,
for which the mean-field diagram is topologically correct
\cite{Shin}. The coefficients determining the second-order phase
transition and the tricritical point are readily calculated as
$\alpha=\partial \omega_{\rm BCS}[0;\mu_\sigma]/\partial
|\Delta|^2$\ and $\beta=\partial^2 \omega_{\rm
BCS}[0;\mu_\sigma]/\partial^2 |\Delta|^2$. The results are shown
in the phase diagram of Fig.\ 1a. The polarization is defined as
$P=(n_+-n_-)/(n_++n_-)$, while the particle densities are
determined by $n_{\sigma}= -\partial\omega_{\rm BCS}[\langle
\Delta \rangle;\mu_\sigma]/\partial \mu_{\sigma}$. Therefore, the
polarization is discontinuous simultaneously with the order
parameter, which gives rise to a forbidden region (FR) below the
tricritical point.

From Fig.\ 1a, we see that the mean-field phase diagram also
contains a Lifshitz point. It is calculated from the
noninteracting Green's function for the Cooper pairs
$G_{\Delta}({\bf k},i \omega_n)$. In the normal state, this
propagator is given by $\hbar G^{-1}_{\Delta}({\bf k},i
\omega_n)=1/T^{2B}-\hbar\Xi({\bf k},i \omega_n)$, where $T^{\rm
2B}=4\pi a\hbar^2/m$ with $a$ the diverging scattering length,
while $\Xi$ is the so-called ladder diagram, given by
\begin{eqnarray}\label{Ladder1}
\hbar \Xi({\bf k},i \omega_n)&=&\int \frac{d{\bf k' }}
{(2\pi)^3}\left\{\frac{1}{2\epsilon({\bf k'})}\right. \\
&+&\left. \frac{1-N_+({\bf k'})-N_-({\bf
k-k'})}{i\hbar\omega_n-\epsilon_{+}({\bf k'})-\epsilon_{-}({\bf
k-k'})+2\mu} \right\} \nonumber
\end{eqnarray}
with $N_{\sigma}({\bf k})=1/[e^{(\epsilon_{\sigma}({\bf k
})-\mu_{\sigma})/k_{\rm B}T}+1]$ the Fermi distributions. Note
that the mean-field expression for $\alpha$ is equal to $-\hbar
G^{-1}_{\Delta}({\bf 0},0)$, while we have that $\gamma=-\partial
\hbar G^{-1}_{\Delta}({\bf 0},0)/\partial {\bf k}^2=0$ at the
Lifshitz point. What precisely happens below the Lifshitz point is
an intriguing question for further research. In Fig. 1a, we have
sketched a simple scenario, where there is a second-order
transition from the normal to the supersolid phase, for which the
condition is $G^{-1}_{\Delta}({\bf k}_{\rm SS},0)=0$ with ${\bf
k}_{\rm SS}$ the wavevector of the supersolid. However, this
transition can in principle be of first order, where the realized
supersolid periodicity can also contain more than one wavevector
\cite{Mora}. Moreover, the transition from supersolidity to the
homogeneous superfluid phase is also expected to be of first
order. The calculation for the stability regions of all possible
supersolid lattices is beyond the scope of this Letter, such that
the dashed lines in Fig.\ 1a are merely guides to the eye.

\begin{figure}[b!]
\includegraphics[width=0.85\columnwidth]{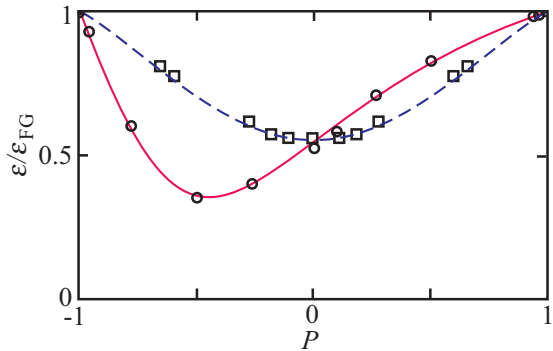}
\caption{\label{radii} (Color online) Equations of state for the
zero-temperature normal state of the ${}^6$Li-${}^{40}$K mixture
(full line) and the mass-balanced mixture (dashed line) in the
unitarity limit. The Monte Carlo results for the
${}^6$Li-${}^{40}$K mixture by Gezerlis {\it et al}.
\cite{Carlson} are shown by circles, while the Monte Carlo results
for the mass-balanced mixture by Lobo {\it et al}. \cite{Lobo} are
shown by squares. The energy density $\epsilon$ is scaled with the
ideal gas result $\epsilon_{\rm FG}= 3(\epsilon_{{\rm F}+} n_{+} +
\epsilon_{{\rm F}-} n_{-})/5$ with $\epsilon_{{\rm
F}\sigma}=\hbar^2(6 \pi^2 n_{\sigma})^{2/3}/2m_{\sigma}$. The
deviations from one thus show the strong interaction effects.}
\end{figure}

\emph{Strong interactions.} --- Having established that within
mean-field theory a tricritical point dominates the phase diagram
when there is an abundance of light ${}^6$Li atoms, while a
Lifshitz points is present when there is an abundance of heavy
${}^{40}$K atoms, the question arises whether this interesting
physics remains when we take fluctuation effects into account.
Answering this question also leads to more quantitative
predictions for future experiments. The normal equation of state
at unitarity is known to be strongly affected by fluctuations
\cite{Melo}. Since this equation of state exactly follows from the
fermionic selfenergies, we first try to take these accurately into
account. We achieve this through a simple construction that gives
excellent agreement with Monte Carlo results, and for which we
only need to know the selfenergy $\hbar\Sigma_{\sigma}$ of a
single $\sigma$ atom in a sea of $-\sigma$ atoms. At zero
temperature, it is given by $\hbar\Sigma_{\sigma}=-c_{\sigma}
\mu_{-\sigma}$. The coefficients can be calculated within the
ladder approximation \cite{Combescot} giving $c_{+}=2.2$ and
$c_{-}=0.44$, or with a renormalization-group approach \cite{Koos}
giving $c_{+}=2.0$ and $c_{-}=0.34$, while Monte Carlo
calculations lead to $c_{+}=2.3$ and $c_{-}=0.36$ \cite{Carlson}.
Noting that these approaches agree rather well, we use the Monte
Carlo results for the rest of the calculation, since they
effectively incorporate all Feynman diagrams. Moreover,
$\hbar\Sigma_{-\sigma}=0$ because the majority atoms are
unaffected by the single minority atom. The chemical potential of
the minority particle is then given by
$\mu_{\sigma}=\hbar\Sigma_{\sigma}$.

Next, we define renormalized chemical potentials as
$\mu'_{\sigma}=\mu_{\sigma}+c_{\sigma}\mu'^2_{-\sigma}/(\mu'_{\sigma}+\mu'_{-\sigma})$
\cite{Diederix}. Two important solutions to these equations are
such that the chemical potential of one species ($-\sigma$) is not
renormalized, $\mu'_{-\sigma}= \mu_{-\sigma}$, while the
renormalized chemical potential of the other species is zero,
i.e., $\mu'_{\sigma}= \mu_{\sigma}+c_{\sigma}\mu_{-\sigma}=0$.
These two solutions correspond precisely to the two extremely
imbalanced limits. By using the renormalized chemical potentials
in the mean-field thermodynamic potential, we can calculate the
full normal equation of state by using the equation for the
densities $n_{\sigma}=-
\partial\omega_{\rm
BCS}[0;\mu'_{\sigma}]/\partial\mu_{\sigma}$. The comparison with
the Monte Carlo equation of state is shown in Fig. 2 and the
agreement is excellent. We also show the comparison for the
mass-balanced case, where both ladder and Monte Carlo calculations
give $c_{\pm}=0.61$. The agreement shows that our construction
captures the full thermodynamics of the strongly interacting
normal state without any free parameters. Since we do not expect
the coefficients $c_{\pm}$ to depend strongly on temperature, we
can also use our method at low temperatures. Moreover, it is
easily extended to other mass ratios and to the superfluid state
\cite{Diederix}.

The above discussion shows that we effectively account for all
fluctuations in the normal state. However, there is a second
effect of particle-hole fluctuations which affects the superfluid
state. Namely, the change in the coefficient $\alpha$ due to
screening of the interspecies interaction \cite{Gorkov}. We
account for this screening by replacing the two-body transition
matrix $T^{\rm 2B}$ with an effective transition matrix that
includes the so-called bubble sum. We thus have that $1/T^{\rm eff
}= 1/T^{\rm 2B}-\hbar \Pi({\bf 0},0)$, where
\begin{eqnarray}\label{Ladder2}
\hbar \Pi({\bf 0},0)&=&\int \frac{d{\bf k' }} {(2\pi)^3}
\frac{N_+({\bf k'})-N_-({\bf k'})}{2h' - \epsilon_{+}({\bf
k'})+\epsilon_{-}({\bf k'})}.
\end{eqnarray}
Since now $\alpha =-\hbar G^{-1}_{\Delta}({\bf 0},0)=-1/T^{\rm eff
}+\hbar\Xi({\bf 0},0)$, the equation for the critical temperature,
$\alpha=0$, becomes $-\hbar \Xi({\bf 0},0)=\hbar \Pi({\bf 0},0)$,
where we use renormalized chemical potentials to include the
selfenergy effects. If we apply this procedure to the mass and
population-balanced case we find $T_{\rm c} = 0.18 T_{\rm F}$ and
$\mu(T_{\rm c})=0.52 \epsilon_{\rm F}$, which is to be compared
with the Monte Carlo results $T_{\rm c} = 0.15 T_{\rm F}$ and
$\mu(T_c)=0.49 \epsilon_{\rm F}$ \cite{Burovski}. Moreover, we
find for the mass-balanced tricritical point that $k_{\rm
B}T_{{\rm c}3} = 0.09 \epsilon_{\rm F+}$ and $P_{{\rm c}3}=0.25$,
which agrees with the experimental data \cite{Shin}. Returning to
the ${}^6$Li-${}^{40}$K mixture and using the same conditions for
the tricritical point and the Lifshitz point as before, we find
the result in Fig. 1b. Note that the critical temperatures are
lowered by about a factor of three due to the screening and the
selfenergy effects. Also the locations of the Lifshitz point and
the tricritical point have drastically changed, since now we find
$T_{{\rm c}3} = 0.08 T_{\rm F}$ and $P_{{\rm c}3} = 0.47$, while
$T_{{\rm L}}=0.05 T_{\rm F }$ and $P_{{\rm L}} =-0.18$.

\emph{Conclusion.}--- We have considered the experimentally
available ${}^6$Li-${}^{40}$K mixture at unitarity, where we
incorporated selfenergy effects to reproduce at zero temperature
the normal equation of state from Monte Carlo calculations. By
also including the effect of screening on the critical
temperature, we have made quantitative predictions for the phase
diagram, which contains both a Lifshitz and a tricritical point.
At weaker attractions the Lifshitz point remains present, although
its temperature gets exponentially suppressed. Below it, various
supersolid phases are competitive \cite{Mora,Yip} and a rich phase
structure is expected. We hope that this Letter will stimulate new
experimental research to supersolidity.

\emph{Acknowledgements.}--- We thank Jeroen Diederix for
stimulating discussions and Alex Gezerlis for kindly providing us
with their Mont Carlo data. This work is supported by the
Stichting voor Fundamenteel Onderzoek der Materie (FOM) and the
Nederlandse Organisatie voor Wetenschaplijk Onderzoek (NWO).

\end{document}